\documentclass[twocolumn,preprintnumbers,superscriptaddress,longbibliography]{revtex4-2}%

\usepackage{braket}
\usepackage{amsmath}
\usepackage{graphicx}
\usepackage{harpoon}

\DeclareMathOperator{\E}{\mathbb{E}}
\DeclareMathAlphabet{\pazocal}{OMS}{zplm}{m}{n}

\usepackage{amssymb}
\usepackage{relsize}
\usepackage{dsfont}

\usepackage{refstyle}
\usepackage{float}
\usepackage{ulem} 

\begin{document}

\title{Protection of quantum information in a chain of Josephson junctions} 

\author{Paul Brookes}
\affiliation{Department of Physics and Astronomy, University College London, London, WC1E 6BT, UK}
\author{Tikai Chang}
\affiliation{Department of Physics and Center for Quantum Entanglement Science and Technology, Bar-Ilan University, 52900 Ramat-Gan, Israel}
\author{Marzena Szymanska}
\affiliation{Department of Physics and Astronomy, University College London, London, WC1E 6BT, UK}
\author{Eytan Grosfeld}
\affiliation{Department of Physics, Ben-Gurion University of the Negev, Beer-Sheva 8410501, Israel}
\author{Eran Ginossar}
\affiliation{Department of Physics and Advanced Technology Institute, University of Surrey, Guildford, GU2 7XH, UK}
\author{Michael Stern}
\thanks{Corresponding author. E-mail: michael.stern@biu.ac.il}
\affiliation{Department of Physics and Center for Quantum Entanglement Science and Technology, Bar-Ilan University, 52900 Ramat-Gan, Israel}

\begin{abstract}
Symmetry considerations are key towards our understanding of the fundamental laws of Nature. The presence of a symmetry implies that a physical system is invariant under specific transformations and this invariance may have deep consequences. For instance, symmetry arguments state that a system will remain in its initial state if incentives to actions are equally balanced. Here, we apply this principle to a chain of qubits and show that it is possible to engineer the symmetries of its Hamiltonian in order to keep quantum information intrinsically protected from both relaxation and decoherence. We show that the coherence properties of this system are strongly enhanced relative to those of its individual components. Such a qubit chain can be realized using a simple architecture consisting of a relatively small number of superconducting Josephson junctions. 
\end{abstract}

\maketitle

\section{Introduction}
One of the greatest obstacles towards the physical realization of a large scale quantum computer with superconducting circuits is the short coherence time of its constituent quantum bits (qubits). During the implementation of an algorithm the state of the qubit register may be rapidly disturbed due to interactions with the environment leading to the accumulation of errors, which prevent the successful completion of the algorithm. Despite significant recent breakthroughs in the development of long lived superconducting qubits \cite{place2021new,somoroff2021millisecond}, high error rates remain a major roadblock in the scaling up of quantum computers and the carrying out of useful calculations. Error correction schemes \cite{kitaev2003fault,dennis2002topological} such as the surface code offer an avenue to circumvent this problem by actively controlling the propagation of errors. However, they are extremely costly, both in terms of the very large number of physical qubits needed to implement a logical fully protected qubit as well as in terms of the amount of classical electronics, computing power, and other overhead resources needed for their implementation.

A possible solution to avoid the complexity of active error correction consists of encoding quantum information in a non-local fashion, so that it is naturally protected by the symmetry ground state properties of a specifically designed Hamiltonian \cite{levitov2001quantum,gladchenko2009superconducting,bell2014protected,callison2017protected}. In this work, we apply this principle to a chain of qubits and show that it is possible to engineer the symmetries of its Hamiltonian to keep quantum information intrinsically protected from both relaxation and decoherence. We show that such a logical qubit  can be realized using a small number of superconducting qubits and study its coherence properties versus local noises.

\section{Spin Model}

We start by presenting the main protection-via-symmetry argument considering a general periodic chain composed of $M$ spins. We introduce the operators $\sigma_m$, $\pazocal{N}$, $\pazocal{T}$ and $\pazocal{I}$. The operator $\sigma_m$ is a local Pauli operator acting on site $m \in \{1,...,M\}$. The operator $\pazocal{N}$ is hermitian and corresponds to the total number of excitations in the chain $\pazocal{N} = \sum_{m=1}^{M} (1 + \sigma^z_m)/2  $. The operator $\pazocal{T}$ is unitary and its action is to displace the chain by a single site $\pazocal{T} \sigma_m \pazocal{T}^{-1} = \sigma_{m+1}$ where $\sigma_{M+1}=\sigma_1$. Finally the operator $\pazocal{I}$ is involutive ($\pazocal{I} = \pazocal{I}^{-1}$) and its action is to invert the chain around a given axis of symmetry $\pazocal{I} \sigma_m \pazocal{I} = \sigma_{M+1-m}$. 

Let us now imagine that we can find two collective states denoted as $\ket{0}$ and $\ket{1}$ which obey the following symmetry relations: 
\begin{align}
& \pazocal{N} \ket{1} = N \ket{1} , \quad   \pazocal{N} \ket{0} = N \ket{0}, \\
& \pazocal{T} \ket{1} = \ket{1} ,  \quad  \quad \pazocal{T} \ket{0} = \ket{0}, \\
& \pazocal{I} \ket{1} = \ket{1}  ,  \quad \quad \pazocal{I} \ket{0} = -\ket{0}.
\end{align}
It is straightforward to show that the action of $U_m = \pazocal{T}^{2m-M-1} \pazocal{I}$ preserves $\sigma_m$ but forces
\begin{align}
\bra{1} \sigma_m \ket{0} = \underbrace{ \bra{1} U_m }_{\bra{1}} \sigma_m \underbrace{ U_m^{-1} \ket{0} }_{-\ket{0}} = - \bra{1} \sigma_m \ket{0} = 0.
\end{align}
Hence the transition matrix element of any single site Pauli operator $\bra{1} \sigma_m \ket{0}$ vanishes. Thus, any noise source coupled locally to any site in the chain via this operator would be unable to cause relaxation between the two states.

\begin{figure*}
\centering
\includegraphics[width=\linewidth]{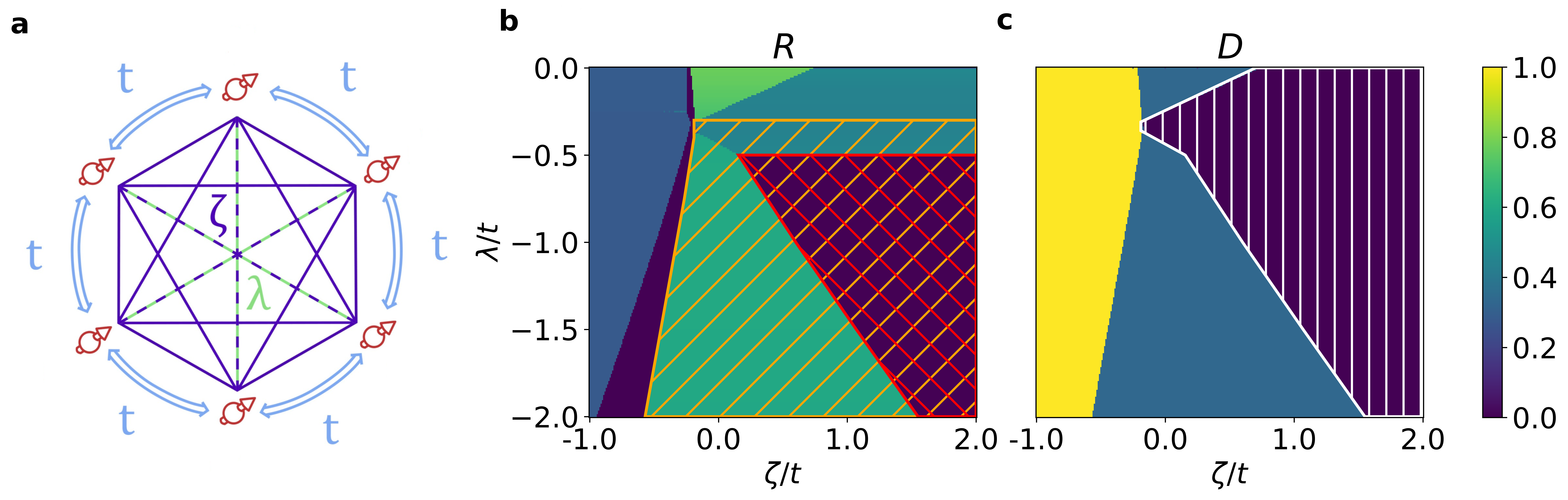}
\caption{\textbf{Decoherence properties of a periodic chain containing six spins with long range couplings.} \textbf{a}, Schematic representation of the Hamiltonian of the chain. Each spin (represented in red) is coupled by flip-flop interactions to both its nearest neighbours ($t$, blue arrows) and to its diametrically opposite counterpart ($\lambda$, green lines). In addition there are $\sigma^z \otimes \sigma^z$ interactions between all spins in the chain ($\zeta$, purple lines). \textbf{b}, Sensitivity to relaxation $R=\sqrt{\sum_{w \in \{x,y,z\}} \vert \bra{1} \sigma^w_m \ket{0} \vert^2}$ vs the strengths of the longer range interactions $\zeta$ and $\lambda$. The red hatched region indicates that both states are invariant under the action of the translation operator $\pazocal{T}$. The orange hatched region indicates that the ground (resp. first excited) state is an eigenvector of the inversion operator $\pazocal{I}$ with eigenvalue +1 (resp. -1). \textbf{c}, Sensitivity to dephasing $D=\sqrt{\sum_{w \in \{x,y,z\}} \vert \bra{1} \sigma^w_m \ket{1} - \bra{0} \sigma^w_m \ket{0} \vert^2}$ vs the strengths of the long range interactions $\zeta$ and $\lambda$. The white vertically hatched region indicates that the ground and first excited states are eigenvectors of the total number of excitations $\pazocal{N}$ with eigenvalue $3$. The dark regions indicate parameters where the ground and first excited state of the chain are protected against local perturbations.}	
\label{fig:1}
\end{figure*}

Furthermore, the system is also protected against local dephasing. Since $[ \pazocal{N}, \sigma^\pm_m ] = \pm \sigma_m^\pm$, the states $\sigma^\pm_m \ket{1}$ and $\sigma^\pm_m \ket{0}$ are eigenstates of $\pazocal{N}$ with eigenvalues $N\pm 1$. Therefore the matrix elements $\bra{1} \sigma^\pm_m \ket{1}$ and $\bra{0} \sigma^\pm_m \ket{0}$ vanish. In addition, we can show that the expectation value of $\pazocal{N}$ for the $\ket{1}$ state is given by:
\begin{align}
    \bra{1} \pazocal{N} \ket{1} &= \frac{1}{2} \sum_{l=0}^{M-1} \bra{1} \pazocal{T}^l (1+\sigma^z_m) \pazocal{T}^{-l} \ket{1} \nonumber \\ &= \frac{M}{2} \bra{1} (1+\sigma^z_m) \ket{1} \text{ for any } m.
\end{align}
Similarly we find $\bra{0} \pazocal{N} \ket{0} = M \bra{0} (1+\sigma^z_m) \ket{0} / 2$. This implies $\bra{1} \sigma^z_m \ket{1} = \bra{0} \sigma^z_m \ket{0} $. Hence, the first order change in the transition energy of our qubit $\bra{1} \sigma_m \ket{1} - \bra{0} \sigma_m \ket{0}$ due to \textit{any local operator} is zero, and so the pure dephasing is cancelled.

The symmetry relations mentioned above have dramatic implications for the coherence of the qubit formed by these two hypothetical states. Unfortunately it is not easy to engineer such a qubit. There are many examples of models which obey the necessary symmetries \cite{majumdar1969next,kitaev2001unpaired,levitov2001quantum,dziarmaga2005dynamics,callison2017protected} and produce this pair of states, yet which cannot be used practically. For example the spectrum of the $XX$ model contains a degenerate manifold, within which lie two states satisfying the required symmetries. However, they cannot be used as a qubit since they are not well isolated from the rest of the spectrum and therefore cannot be easily initialised and controlled.

\begin{figure*}[t]
\includegraphics[width=\linewidth]{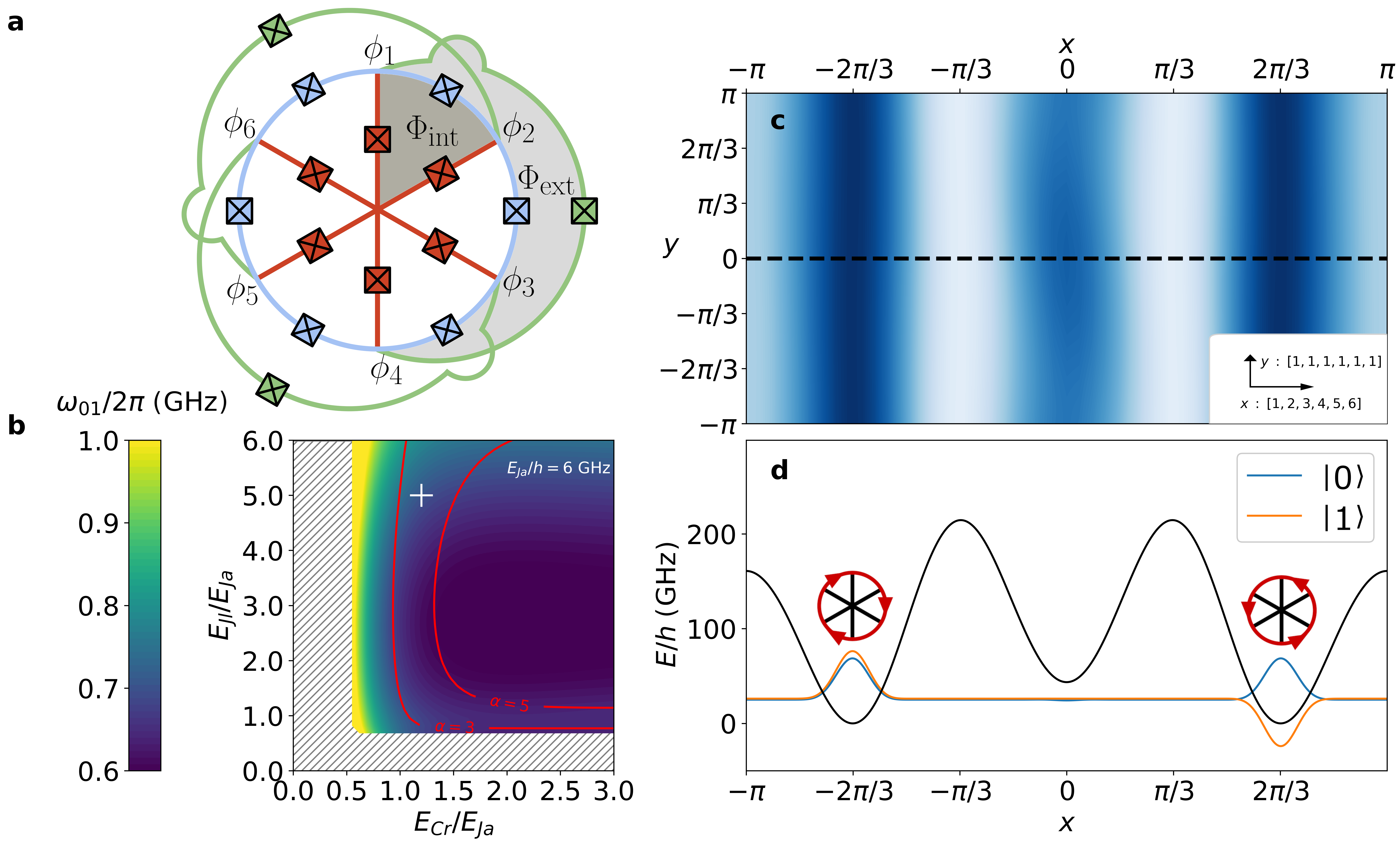}
\caption{\textbf{Realizing the protected states in a superconducting circuit: the Magenium qubit.} \textbf{a}, Circuit diagram of the qubit showing a superconducting loop intersected by six identical azimuthal Josephson junctions of Josephson energy $E_{Ja}$ (in blue). Each of the islands - whose phases are labelled $\phi_1$ to $\phi_6$ - is connected both radially to ground via a Josephson junction of charging energy $E_{Cr}$ (in red) and to its diametrically opposite counterpart by a junction with Josephson energy $E_{Jl}$ (in green). \textbf{b}, Transition frequency to the first excited state of the qubit $\omega_{01}/2\pi$ as a function of $E_{Cr}/E_{Ja}$ and $E_{Jl}/E_{Ja}$ assuming all the junctions of the circuit have the same bare plasma frequency $\sqrt{8 E_J E_C} / h = 10 ~ \mathrm{GHz}$. The hatched region corresponds to parameters where the ground and first excited states do not exhibit the required symmetries for protection of the qubit against decoherence. The red lines indicate contours of constant anharmonicity $\alpha = \omega_{12}/\omega_{01}$. The white cross corresponds to the qubit parameters we chose for the remainder of the article. \textbf{c}, Potential landscape in the $(x,y)$ plane. In this plane two valleys of global minima can be seen. The wavefunctions of the ground and first excited state are localised in the potential minima.  \textbf{d}, Potential energy along the horizontal direction shown in \textbf{c} as a dashed black line. Two global potential minima are observed for $x=\pm 2 \pi / 3$ and correspond to states with clockwise or anti-clockwise current flowing in the circuit. The ground (resp. first excited) state wavefunction consists of a symmetric (resp. anti-symmetric) superposition of these states.}\label{fig:2}
\end{figure*}

In this work we propose a physical system where such states can be observed and efficiently used as a qubit. To achieve this we first consider a chain formed by six spins on a ring. The Hamiltonian of this system can be written as
\begin{align}
\label{eq:ham_spin_half_chain}
    H =& \frac{\zeta}{4} \sum_{m,n=1}^{M} \sigma_m^z \sigma_{n}^z  + \frac{t}{2} \sum_{m=1}^{M}  (\sigma_m^+ \sigma_{m+1}^- + \sigma_{m}^- \sigma_{m+1}^+) \nonumber \\ & + \frac{\lambda}{2} \sum_{m=1}^{M} (\sigma_m^+ \sigma_{m+3}^- + \sigma_{m}^- \sigma_{m+3}^+)
\end{align}
in which flip-flop interactions exchange excitations between neighbouring sites with a coupling strength $t$ and between diametrically opposite sites with a coupling strength $\lambda$. In addition there is an all-to-all $\sigma^z \otimes \sigma^z$ coupling with strength $\zeta$. In Fig. \ref{fig:1}a we represent this model diagrammatically and show in Fig. \ref{fig:1}b and c that it is possible to produce a qubit manifold which satisfies the above symmetries and is well separated from the rest of the spectrum. Within the parameter space where the symmetries are obeyed (indicated by hatched regions), we indeed observe that the relaxation and dephasing rate of the qubit are suppressed! However we notice that the strength of the long range interactions must be of similar strength to the nearest neighbour interactions and of opposite sign.

\section{The Magenium Qubit}

The Magenium qubit is an implementation of such a system using an array of Josephson junctions and the toolset of circuit-QED. In the methods section we demonstrate the mapping between the spin Hamiltonian above and the full Hamiltonian derived from rigorous circuit analysis. In the Magenium, the spins are formed by the radial Josephson junctions shown in red in Fig. \ref{fig:2}a, which have Josephson energy $E_{Jr}$ and charging energy $E_{Cr}$. Each junction forms a Cooper pair box superconducting qubit and is tuned to a gate charge of half a Cooper pair ($N_g = 1/2$). These qubits are coupled by flip-flop interactions to their nearest neighbours via azimuthal Josephson junctions with Josephson energy $E_{Ja}$ and to their diametrically opposite counterparts via Josephson junctions of Josephson energy $E_{Jl}$. All subsequent calculations are carried out using the full circuit model and including 10 charge states on each of the six nodes of the circuit. We choose all three junction types to have the same bare plasma frequency. This frequency is set to $\sqrt{8 E_J E_C} / h = 10 ~ \mathrm{GHz}$, which is typical for static oxidation of transmon junctions. The charging energies of the azimuthal junctions $E_{Ca}$ are chosen such that $E_{Ca}/E_{Cr} \ll 1$. In this case a long range charge coupling of strength $\sim E_{Cr}$ arises between the qubits which gives rise to an all-to-all $\sigma^z \otimes \sigma^z$ coupling.

The relative sign of the flip-flop interactions is controlled by the choice of the flux threading through the circuit. At the operating point, the flux threading through each internal sector is tuned to half a flux quantum $\Phi_{\mathrm{int}} = \Phi_0 / 2$, while the flux threading the outer loops is $\Phi_{\mathrm{ext}} = 3 \Phi_0 / 2$. This can be easily arranged by applying a homogeneous external magnetic field and choosing the appropriate circuit geometry. In Fig. \ref{fig:2}b we plot the transition frequency between the ground ($\ket{0}$) and first excited ($\ket{1}$) states of the circuit $\omega_{01}/2\pi$ as a function of $E_{Cr}/E_{Ja}$ and $E_{Jl}/E_{Ja}$ assuming $E_{Ja} / h = 6 ~ \mathrm{GHz}$. These axes are analogous to those used in Fig. \ref{fig:1}b since the strengths of the nearest neighbour flip-flop, range 3 flip-flop and all-to-all $\sigma^z \otimes \sigma^z$ couplings are proportional to $E_{Ja}$, $E_{Jl}$ and $E_{Cr}$ respectively.

In the plotted region these states satisfy the desired symmetries. We identify a particular set of parameters ($E_{Jr}/h = 1.7 ~\mathrm{GHz}$, $E_{Cr}/h = 7.4 ~\mathrm{GHz}$, $E_{Ja}/h = 6.0 ~\mathrm{GHz}$, $E_{Ca}/h = 2.1 ~\mathrm{GHz}$, $E_{Jl}/h = 30.0 ~\mathrm{GHz}$ and $E_{Cl}/h = 0.56 ~\mathrm{GHz}$), marked by the white cross, which will be used in the following to demonstrate our design. At this location the transition frequency of the qubit is $704 ~\mathrm{MHz}$ and the anharmonicity of transitions to the next excited state is $\omega_{12}/\omega_{01} = 4$, where $\omega_{12}$ is the transition frequency between the first and second excited states of the circuit. Given this large anharmonicity we will be able to address our qubit states without the participation of higher levels.

In order to better understand the nature of the $\ket{1}$ and $\ket{0}$ states we represent the potential formed by the junctions of the circuit as a function of the six node phases $(\phi_1,\phi_2,\phi_3,\phi_4,\phi_5,\phi_6)$. We observe that the global minima of the potential lie within a plane. Within this plane the node phases are given by $\phi_n = n x + y$. We observe that the potential minima form two valleys along the lines $y = x \pm \frac{2\pi}{3}$ (Fig. \ref{fig:2}c).

In Fig. \ref{fig:2}d we plot the potential along $y=0$ and the wavefunctions of the qubit states projected along this axis. We observe that the wavefunctions are localised in the potential minima, which are located at $x = \pm 2 \pi / 3$. The co-ordinates of the left minimum $\overline{\phi}_\circlearrowright = (-2\pi/3,-4\pi/3,-6\pi/3,-8\pi/3,-10\pi/3,-12\pi/3)$ correspond to a clockwise persistent current flowing in each azimuthal junction $I_p = I_{ca} \sin(\pi/3)$ where $I_{ca}$ is the critical current of these junctions, while the current in the junctions of the outer loops are all zero. Similarly, the co-ordinates of the right minimum $\overline{\phi}_\circlearrowleft = (2\pi/3,4\pi/3,6\pi/3,8\pi/3,10\pi/3,12\pi/3)$ correspond to an anti-clockwise persistent current $I_p = I_{ca} \sin(\pi/3)$.
We denote the states localised in these minima by $\ket{\circlearrowright}$ and $\ket{\circlearrowleft}$ so that the qubit states are given by $\frac{1}{\sqrt{2}}(\ket{\circlearrowright} \pm \ket{\circlearrowleft})$. The presence of the valley along $y$ testifies that these circulating states are eigenstates of the total excitation operator $\pazocal{N} = i \partial_y$.

\section{Coherence Properties}

\begin{figure}
\centering
\includegraphics[width=\linewidth]{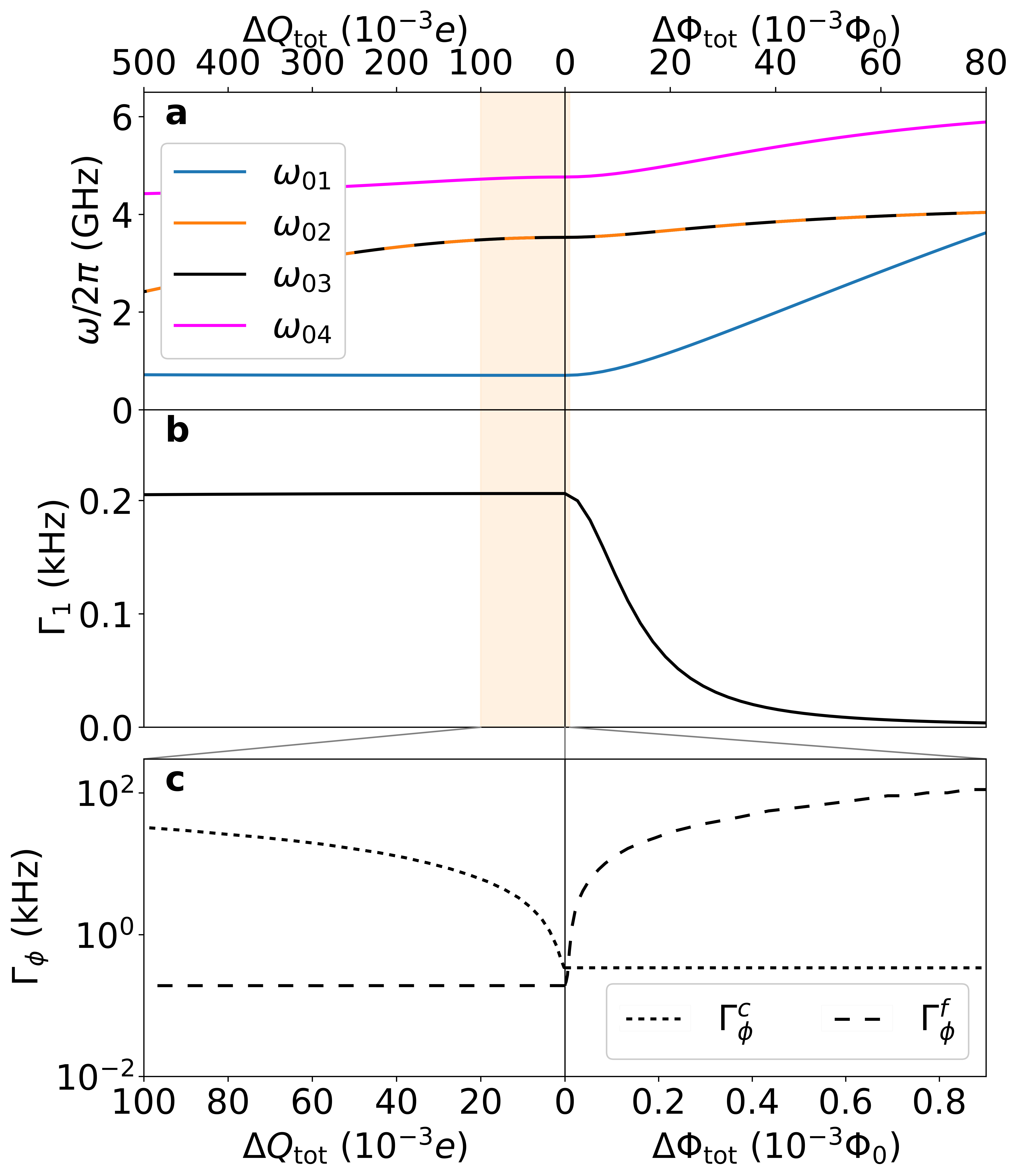}
\caption{\textbf{Decoherence properties of the Magenium qubit}. \textbf{a}, Spectroscopy of the Magenium qubit showing the first four transition frequencies vs. the total charge $\Delta Q_{\mathrm{tot}}$ and flux $\Delta \Phi_{\mathrm{tot}}$ offsets from the operating point. \textbf{b}, Relaxation rate of the Magenium qubit due to quasiparticle tunneling assuming a quasiparticle density of  $n_\mathrm{qp}=0.05 / \mu \mathrm{m}^{-3}$. \textbf{c}, Pure dephasing rate away from the optimal point due to charge noise ($\Gamma_\phi^c$) or flux noise ($\Gamma_\phi^f$). The spectral density of the charge and flux noise is assumed to behave as $1/f$ with an IR cutoff of $1 ~\mathrm{Hz}$ and a UV cutoff of $1 ~\mathrm{MHz}$. The total strengths of the power spectra at $1 ~\mathrm{Hz}$ are assumed to be $A^c = (2 \times 10^{-4} e)^2$ for charge noise and $A^f = (2 \times 10^{-6} \Phi_0)^2$ for flux noise. The noise is divided evenly among the gates and loops of the circuit, which are considered to be independent sources of noise.}\label{fig:4}
\end{figure}

In Fig. \ref{fig:4}a we plot how the energies of the four lowest excited states of the Hamiltonian change as we vary the total gate charge and flux offsets: $\Delta Q_{\mathrm{tot}}$ and $\Delta \Phi_{\mathrm{tot}}$. Varying the gate charges does not have a significant effect on the transition frequency of the qubit, but varying the flux away from the operating point generates a magnetic dipole where one current state ($\ket{\circlearrowleft}$ or $\ket{\circlearrowright}$) becomes more favourable and therefore increases the transition frequency by $\sim I_p \Delta \Phi_{\mathrm{tot}} / h$. This behaviour is similar to what is usually observed with flux qubits \cite{yoshihara2006decoherence,manucharyan2009fluxonium,bylander2011noise,stern2014flux,somoroff2021millisecond} but the sensitivity of the transition frequency to changes in the flux is significantly reduced since the persistent current flowing in our circuit $I_p = 10~\mathrm{nA}$ is almost two orders of magnitude smaller than the typical current flowing in a flux qubit. Eventually the transition frequency will increase until the first and second excited states cross at $\omega_{01}/2\pi = 4 ~\mathrm{GHz}$.

We now turn our attention to the coherence properties of the Magenium qubit both at the operating point $(\Phi_{\mathrm{int}} = \Phi_0 / 2, \Phi_{\mathrm{ext}} = 3 \Phi_0 / 2, N_g = 1/2 )$ and away from it. Relaxation via dielectric dissipation normally occurs due to coupling of eigenstates via the number operators $\hat{N}_m$, \cite{stern2014flux} but in our case these vanish and thus this form of relaxation is suppressed. On the other hand, relaxation via quasiparticle tunneling \cite{lutchyn2005quasiparticle,catelani2011quasiparticle,catelani2011relaxation,stern2014flux} couples two sites of the chain and thus is not intrinsically protected by our symmetry arguments. The estimated rate of relaxation due to quasiparticle tunneling is plotted in Fig. \ref{fig:4}b. This rate is proportional to the density of quasiparticles in the circuit. This density is not thermal but is mainly formed by pair breaking photons due to stray radiation \cite{glazman2021bogoliubov,serniak2018hot} and thus can be reduced by appropriate filtering \cite{serniak2019direct}.  Assuming a density of $n_\mathrm{qp} = 0.05 ~ \mu \mathrm{m}^{-3}$, we find a relaxation rate of $\Gamma_1 = 0.2 ~ \mathrm{kHz}$, which corresponds to a relaxation time of $\, T_1 = 5 ~ \mathrm{ms}$. The relaxation rate is insensitive to detuning of the charge from the optimal point, but it decreases when the flux is detuned.

In addition, we examine how coherence properties of the qubit are limited by higher order effects of dephasing channels such as flux noise and charge noise. In Fig. \ref{fig:4}c we plot the rate of dephasing up to second order in both charge and flux noise. We assume that these noise sources have a $1/f$ spectrum with an IR cutoff of $1 ~\mathrm{Hz}$ and a UV cutoff of $1 ~\mathrm{MHz}$. The total strengths of these power spectra at $1 ~\mathrm{Hz}$ are assumed to be $A_c = (2 \times 10^{-4} e)^2$ for charge noise \cite{kuzmin1989single,geerligs1990tunneling,zimmerli1992noise,wolf1997investigation,martinis2003decoherence} and $A_f = (2 \times 10^{-6} \Phi_0)^2$ for flux noise \cite{yoshihara2006decoherence,stern2014flux,braumuller2020characterizing}. The noise power is divided evenly among the gate charges and loops of the circuit, each of which are treated as independent sources of noise. At the optimal point the charge and flux noises give pure dephasing times of $2.9 ~ \mathrm{ms}$ and $5.2 ~ \mathrm{ms}$ respectively. Moreover, charge noise may also include rare events where a single electron will suddenly appear on one of the islands of the qubit \cite{wills2021characterisation}. To handle this issue, the center island of the qubit will be engineered to act as a quasiparticle trap \cite{sun2012measurements}.

These coherence times are promising, but we must be careful to also consider the effect of disorder, which will unavoidably occur in practice. In particular we expect there to be some variation in the size of the junctions, in their oxidation parameters, in the areas of the loops and in the gate charges on each island. To take this disorder into account we introduce variations in the tunneling energies, charging energies and loop fluxes of the form $E_J \rightarrow E_J (1 + X_m) (1+ Y_m)$, $E_C \rightarrow E_C / (1 + X_m)$ and $\Phi_{\mathrm{int(ext)}} \rightarrow \Phi_{\mathrm{int(ext)}} (1 + Z_m) $ where $X_m$, $Y_m$ and $Z_m$ are independent random variables sampled from a normal distribution with standard deviation $\sigma = 2 \%$ for $X_m$ and $Y_m$, and $\sigma = 0.2 \%$ for $Z_m$. Remarkably, this disorder slightly shifts the qubit frequency by $\approx 20 ~ \mathrm{MHz}$ and shifts the magnetic field strength of the operating point but otherwise leaves the coherence time unchanged. On the other hand, the limits on gate charge disorder are more stringent: a level of disorder in the gate charges of $0.1 \%$ leaves the qubit frequency unchanged but reduces the coherence time to $\approx 1.4 ~\mathrm{ms}$. Electrostatic simulations indicate this level of disorder is attainable using a single back gate \cite{supplementary}. An experimental realisation of the Magenium qubit is under investigation and will be the subject of a forthcoming publication.

\hspace{0.5cm}

\section*{Acknowledgements}
This work is supported by the Israeli Science Foundation under grant numbers 1626/16 and 898/19, and by the EPSRC under grant numbers EP/T001062/1, EP/K003623/2 and EP/S019669/1. E. Ginossar and M. Stern acknowledge support from the International Exchange awards of the Royal Society under project entitled "Long range interactions in superconducting circuits". P. Brookes acknowledges support from the UCLQ Challenge Innovation Fund. M. Stern wishes to thank G. Catelani for fruitful discussions on dephasing due to quasiparticle tunneling in the device.

\section{Methods}

\subsection*{Hamiltonian of the Magenium Qubit}

We now outline the process of quantizing our circuit to obtain its Hamiltonian. This process is explained in more detail in \cite{devoret1995quantum,burkard2004multilevel,bishop2010circuit}. The circuit is treated as a graph consisting of a grounded node in the centre connected to 6 outer nodes via radial Josephson junctions. The voltage of node $n$ at time $t$ is written as $v_n(t)$, from which we define the node fluxes by $\phi_n(t)  = \int_{-\infty}^{t} v_n(t^\prime) d t^\prime$. The Lagrangian $L$ is then divided into a kinetic part $T$ which consists of the capacitive charging energies, and a potential part $V$ which is the sum of the inductive energies of the Josephson junctions: $L=T-V$. If we denote the capacitances of the radial, azimuthal (nearest-neighbour) and outer (next-next-nearest-neighbour) junctions by $C_r$, $C_a$ and $C_l$ respectively then we can write the kinetic term as
\begin{align}
    T =& \frac{1}{2} \sum_{m=1}^{6} \big[ C_a (\dot{\phi}_{m+1}-\dot{\phi}_{m})^2 + C_r \dot{\phi}_{m}^2 \big] \nonumber \\ &+ \frac{1}{2} \sum_{m=1}^{3} C_l (\dot{\phi}_{2m}-\dot{\phi}_{2m+3})^2 \\
    =& \frac{1}{2} \sum_{m,n=1}^{6} \mathbf{C}_{mn} \dot{\phi}_m \dot{\phi}_n
\end{align}
where in the second line we have introduced the capacitance matrix:

\begin{widetext}
\begin{align}
\mathbf{C} =
\begin{pmatrix}
C_r \!+\! 2 C_a \!+\! C_l \!\!\!\!\!\!\!\!\! & -C_a & 0 & -C_l & 0 & -C_a \\
-C_a & C_r \!+\! 2 C_a \!+\! C_l \!\!\!\!\!\!\!\!\! & -C_a & 0 & -C_l & 0 \\
0 & -C_a & C_r \!+\! 2 C_a \!+\! C_l \!\!\!\!\!\!\!\!\! & -C_a & 0 & -C_l \\
-C_l & 0 & -C_a & C_r \!+\! 2 C_a \!+\! C_l \!\!\!\!\!\!\!\!\! & -C_a & 0 \\
0 & -C_l & 0 & -C_a & C_r \!+\! 2 C_a \!+\! C_l \!\!\!\!\!\!\!\!\! & -C_a \\
-C_a & 0 & -C_l & 0 & -C_a & C_r \!+\! 2 C_a \!+\! C_l
\end{pmatrix}.
\end{align}
\end{widetext}

In order to write down the potential term we must be careful to take account of any fluxes which may be threading through the loops of the circuit. We first define a spanning tree which reaches all nodes of the circuit without forming any loops. In our case this spanning tree consists of all the radial Josephson junctions. Each Josephson junction in the circuit makes a contribution to the inductive energy of the form $- E_J \cos(\phi / \varphi_0)$ where $\phi$ is the difference in node fluxes across that junction and $\varphi_0$ is the reduced flux quantum. For junctions lying within the spanning tree, i.e. the radial junctions, this phase difference is simply given by the node fluxes $\phi_n(t)$. However if the junction lies outside the spanning tree then the flux difference must account for the external flux threaded through the loop it forms. According to Maxwell's equations the change in potential energy when traversing a loop is proportional to the rate of change of magnetic flux through that loop:
\begin{align}
    \oint \mathbf{E} \cdot d \mathbf{x} = - \partial_t \Phi_{\text{ext}}.
\end{align}
If we integrate this relation over time then we find that the sum of flux differences across circuit elements within a loop will be equal to the external flux threading the loop. This allows us to write the flux differences across the $m$th azimuthal $\Delta \phi_{a,m}$ and outer $\Delta \phi_{l,m}$ junctions as
\begin{align}
    \Delta \phi_{a,m} &= \phi_{m+1} - \phi_{m} - \Phi_{a,m} \\
    \Delta \phi_{l,m} &= \phi_{2m+3} - \phi_{2m} - \Phi_{l,m}
\end{align}
where $\Phi_{a,m}$ and $\Phi_{l,m}$ are the external fluxes threading through the loops formed by these junctions. Finally, these flux differences can be used to write the potential part of the Lagrangian as
\begin{align}
    V =  &- E_{Ja} \sum_{m=1}^{6} \cos \bigg( \frac{\phi_{m+1} - \phi_m - \Phi_{am}}{\varphi_0} \bigg) \nonumber \\ &- E_{Jl} \sum_{m=1}^{3} \cos \bigg( \frac{\phi_{2m+3} - \phi_{2m} - \Phi_{lm}}{\varphi_0} \bigg) \nonumber \\  &- E_{Jr} \sum_{m=1}^{6} \cos \bigg( \frac{\phi_m}{\varphi_0} \bigg).
\end{align}

To convert this Lagrangian to the form of a Hamiltonian we must first obtain the node charges. These are given by
\begin{align}
    q_m &= \frac{\partial L}{\partial \dot{\phi}_m} = \sum_{n=1}^6 C_{mn} \dot{\phi}_n
\end{align}
After performing a Legendre transformation the Hamiltonian is given by
\begin{align}
    H &= \sum_{m=1}^6 \dot{\phi}_m q_m - L = T + V
\end{align}
where we now express the kinetic term as
\begin{align}
T = \frac{1}{2} \sum_{m,n=1}^{6} \mathbf{C}_{mn}^{-1} q_m q_n
\end{align}
This Hamiltonian can be quantized by replacing $q_n \rightarrow 2e (\hat{N}_n - N_{g,n})$ and $\phi_n / \varphi_0 \rightarrow \hat{\theta}_n$ where $N_{g,n}$ is the gate charge on site $n$ and the commutation relation $[\hat{\theta}_n, \hat{N}_n] = i$ holds. This gives the complete Hamiltonian in the form
\begin{align}
\label{eq:circuit_hamiltonian}
    H =& 2 e^2 \sum_{m,n=1}^{6} \mathbf{C}_{mn}^{-1} (\hat{N}_m - N_{g,m}) (\hat{N}_n - N_{g,n}) \nonumber \\ &- E_{Jr} \sum_{m=1}^{6} \cos \bigg( \hat{\theta}_m \bigg) \nonumber \\
    &- E_{Ja} \sum_{m=1}^{6} \cos \bigg( \hat{\theta}_{m+1} - \hat{\theta}_m - \frac{\Phi_{am}}{\varphi_0} \bigg) \nonumber  \\ 
    & - E_{Jl} \sum_{m=1}^{3} \cos \bigg( \hat{\theta}_{2m+3} - \hat{\theta}_{2m} - \frac{\Phi_{lm}}{\varphi_0} \bigg).
\end{align}
If we denote the fluxes through the six inner and three outer loops of the circuit by $\Phi_{I,m}$ and $\Phi_{O,m}$ then we can rewrite the external fluxes as:
\begin{align}
    \Phi_{am} &= \Phi_{I,m}, \nonumber \\ \Phi_{lm} &= \Phi_{I,m} + \Phi_{I,m+1} + \Phi_{I,m+2} + \Phi_{O,m}.
\end{align}
Using this circuit Hamiltonian we are able to numerically calculate the eigenstates of the Magenium qubit and study their response to noise and disorder, as explained in subsequent sections and referenced in the main text.

In addition we can now demonstrate the relationship between this Hamiltonian and our original spin model. We start with the potential term which recreates the flip-flop couplings of the spin-1/2 model. We can see this by rewriting the potential in terms of the tunneling operators
\begin{align}
    \Sigma_m^+ &= \sum_n \ket{n+1}_m \! \bra{n}_m, \\
    \Sigma_m^- &= \sum_n \ket{n}_m \! \bra{n+1}_m
\end{align}
which  cause Cooper pairs to tunnel back and forth across the radial junctions, and are written in terms of the Cooper pair number states $\hat{N}_m \ket{n}_m = n \ket{n}_m$. These operators allow us to represent the cosine and sine functions as
\begin{align}
    \cos(\hat{\theta}_m) &= \frac{1}{2} \Big( \Sigma^-_m + \Sigma^+_m \Big), \\
    \sin(\hat{\theta}_m) &= \frac{i}{2} \Big( \Sigma^-_m - \Sigma^+_m \Big).
\end{align}
Using compound angle formulae the nearest neighbour coupling is then rewritten as
\begin{align}
    \cos & \bigg( \hat{\theta}_{m+1} - \hat{\theta}_m - \frac{\Phi_{am}}{\varphi_0} \bigg) = \nonumber \\ & \frac{1}{2} \bigg( \Sigma^+_m \Sigma^-_{m+1} e^{i \Phi_{am} / \varphi_0} + \Sigma^-_m \Sigma^+_{m+1} e^{-i \Phi_{am} / \varphi_0} \bigg).
\end{align}
Similarly the diametric coupling is rewritten as
\begin{align}
    \cos & \bigg( \hat{\theta}_{m+3} - \hat{\theta}_m - \frac{\Phi_{lm}}{\varphi_0} \bigg) = \nonumber \\ & \frac{1}{2} \bigg( \Sigma^+_m \Sigma^-_{m+3} e^{i \Phi_{lm} / \varphi_0} + \Sigma^-_m \Sigma^+_{m+3} e^{-i \Phi_{lm} / \varphi_0} \bigg).
\end{align}
In order to arrange the current signs for these couplings we simply choose $\Phi_{am}/\varphi_0 = \pi$ and $\Phi_{lm}/\varphi_0 = 4\pi$.

Next we look at the charge coupling. In the limit $C_r / C_a \rightarrow 0$ the inverse of the capacitance matrix gives an all to all charge coupling according to
\begin{align}
    \mathbf{C}^{-1}_{mn} \rightarrow \frac{1}{6 C_r}, \quad\quad T \rightarrow \frac{2}{3} E_{Cr} \sum_{m,n=1}^{6} (\hat{N}_m - N_{g,m})(\hat{N}_n - N_{g,n}).
\end{align}
If the gate charges are tuned to a half integer and we truncate down to two charge states per node then we can make the identification $\hat{N} - N_{g} \sim \frac{1}{2} \sigma^z$ and this coupling will be of the same form as the all-to-all $\sigma^z_m \sigma^z_n$ coupling.

In this manner we can use our circuit to engineer a Hamiltonian which is analogous to the simple spin-1/2 model with Cooper pairs now taking the role of the excitations which can now tunnel between sites via Josephson junctions. Our circuit respects the key translation and inversion symmetries we identified earlier however whereas this previous model consisted of two level sites, each site in our circuit has many levels.

\subsection*{Dephasing}

In order to model the dephasing time of the Magenium qubit we treat the noise as an adiabatically varying parameter in the  Hamiltonian and study its effect on the frequency of the qubit \cite{ithier2005manipulation,ithier2005decoherence}. We consider a Hamiltonian $H(\vec{x})$ which depends on some time varying set of noisy parameters $\vec{x}(t)$. If the power spectral density (PSD) of the noise is limited to frequencies below the gaps between the eigenstates of the Hamiltonian then it will be unable to cause transitions. Therefore we can treat the noise adiabatically and consider the occupation probabilities of all eigenstates to be constant. Our qubit is formed by the lowest two eigenstates of $H\big( \vec{x}(t) \big)$, whose splitting frequency can be treated as a function of $\vec{x}(t)$ and is written as
\begin{align}
\omega_{01}(\vec{x}(t)) = \omega_{01} \big( \vec{x}(0) \big) + \delta \! \omega_{01} \big(\vec{x}(t) \big)
\end{align}
where $\delta \! \omega_{01}(\vec{x}(t))$ denotes any fluctuations in the splitting frequency away from its initial value due to the noise. The state of our qubit consists of a superposition of these states and is written as
\begin{align}
\ket{\psi(t)} = \alpha(t) \ket{0} + \beta(t) \ket{1}.
\end{align}
In order to study the coherence of the qubit during its evolution we must calculate the ensemble average state over many experiments. This will be given by
\begin{align}
\rho(t) &= \E [ \ket{\psi(t)} \bra{\psi(t)} ] \\
&=
\begin{pmatrix}
p_0 & C(t) \\ 
C^{*}(t) & p_1
\end{pmatrix}
\end{align}
where $p_0$ and $p_1$ represent the occupation probabilities of states $\ket{0}$ and $\ket{1}$ and $C(t)$ represents the coherence of their superposition. The evolution of this coherence can be written as
\begin{align} \label{eq:coherence_vs_time}
C(t) &= \E[\alpha(t) \beta^*(t)], \nonumber \\
&= \sqrt{p_0 p_1} \exp \Big(-i \omega_{01} \big(\vec{x}(0) \big) t \Big) f_\phi(t)
\end{align}
where $f_\phi(t) = \E [\exp(- i \phi(t))]$ and $\phi(t) = \int_{0}^{t} \delta \! \omega_{01}\big(\vec{x}(\tau)\big) d\tau$ \cite{ithier2005manipulation,ithier2005decoherence}. This evolution consists of two principal parts: an oscillation at the average frequency of the qubit multiplied by a term which accounts for the variations in the accumulated phase in each experimental run. The latter term decays over time due to destructive interference between these variations, causing the qubit to decohere. Our task is to calculate $f_\phi(t)$ numerically by taking many samples of the noise from an appropriate PSD, calculating and integrating $\delta\omega_{01}(t)$ to produce $\phi(t)$ for each sample, and finally taking the ensemble average of the phase. The dephasing time is the time over which $f_\phi$ decays from $1$ to $1/e$.

We can express the variation in the qubit frequency $\delta\omega_{01}(t)$ due to the noise $\vec{x}(t)$ using a Taylor expansion as
\begin{align} \label{eq:gap_taylor_expansion}
\delta \! \omega_{01}(t) =& \nonumber \\  \sum_m x_m(t) & D_{x_m} + \frac{1}{2} \sum_{m,n} D_{x_m,x_n} x_m(t) x_n(t) + \pazocal{O}(x^3(t))
\end{align}
for which we define
\begin{equation}
D_{x_m} = \partial_{x_m} \omega_{01}
\quad\text{and}\quad 
D_{x_m,x_n} = \partial_{x_m} \partial_{x_n} \omega_{01}.
\end{equation}
These derivatives can be calculated numerically by diagonalizing $H(\vec{x})$ for many samples of $\vec{x}$ and fitting a Taylor expansion to the resulting values of $\omega_{01}(\vec{x})$.

A realization of the noise can be constructed by applying an inverse discrete Fourier transform to components sampled from the PSD, as described in \cite{timmer1995generating}. The PSD of the noise variable $x_m(t)$ is defined by
\begin{align}
S_{x_m}(f) &= \int_{-\infty}^{+\infty} R_{x_m}(\tau) \exp(-2 \pi i f \tau) d\tau, \\
R_{x_m}(\tau) &= \E [ x_m(\tau) x_m(t+\tau) ].
\end{align}
We divide the PSD into $N$ bins of width $\Delta f$, each of which corresponds to a Fourier component in the signal. The amplitude of each component is generated according to
\begin{align}
    \tilde{X}_{m,k} = Z_{m,k} \sqrt{S_{x_m}(f_k) \Delta f}
\end{align}
where $f_k = k \Delta f$ and $Z_{m,k}$ is a complex random variable
\begin{align}
    Z_{m,k} \sim \frac{1}{\sqrt{2}} \bigg( \mathcal{N}(0,1) + i \mathcal{N}(0,1) \bigg)
\end{align}
in which $\mathcal{N}(\mu,\sigma^2)$ is a normal distribution with mean $\mu$ and variance $\sigma^2$. This allows us to construct the sampled signal according to
\begin{align}
    x_m(t_l) = \sum_{k= \frac{1-N}{2}}^{\frac{N-1}{2}} \tilde{X}_{m,k} \exp(2 \pi i f_k t) \label{eq:noise_sample}
\end{align}
in which the time point is given by $t_l = l \Delta t$ and the timestep is given by $\Delta t = \frac{1}{N \Delta f}$. The numbers of samples $N$ is assumed to be odd in the above. In order to produce a real noise signal we must impose $Z_{m,k} = Z_{m,-k}^*$.

We can now combine Eqs. (\ref{eq:gap_taylor_expansion}) and (\ref{eq:noise_sample}) to numerically estimate the time dependence of $\delta \omega_{01}(t)$ and calculate the dephasing time according to Eq. (\ref{eq:coherence_vs_time}). The dephasing calculations in the main text use a PSD of the form
\begin{equation}
S_{x_m}(f) =
  \begin{cases}
    A & \text{for } \vert f \vert < f_\mathrm{IR} \\
    \frac{A}{f} & \text{for } f_\mathrm{IR} \leq \vert f \vert < f_\mathrm{UV} \\
    0 & \text{for } f_\mathrm{UV} \leq \vert f \vert
  \end{cases}
\end{equation}
In the main body of the text we sample from this distribution using $N = 2 \times 10^6 - 1$, $f_\mathrm{IR} = 1 ~\mathrm{Hz}$, $f_\mathrm{UV} = 1.0 ~\mathrm{MHz}$ and $\Delta f = 0.5 ~\mathrm{Hz}$.

\newpage

\bibliography{magenium_bibliography.bib}

\end{document}